\documentclass{article}
\usepackage{amsmath, amssymb, amsfonts}
\usepackage{epsfig}
\title{On a regular charged black hole with a nonlinear electric source}  
\author{Hristu Culetu, \\Ovidius University, Dept.of Physics, \\ Mamaia Avenue 124, 900527 Constanta, Romania, \\e-mail : hculetu@yahoo.com}

\begin{document}
\numberwithin{equation}{section}
\pagenumbering{arabic}
\maketitle
\newcommand{\fv}{\boldsymbol{f}}
\newcommand{\tv}{\boldsymbol{t}}
\newcommand{\gv}{\boldsymbol{g}}
\newcommand{\OV}{\boldsymbol{O}}
\newcommand{\wv}{\boldsymbol{w}}
\newcommand{\WV}{\boldsymbol{W}}
\newcommand{\NV}{\boldsymbol{N}}
\newcommand{\hv}{\boldsymbol{h}}
\newcommand{\yv}{\boldsymbol{y}}
\newcommand{\RE}{\textrm{Re}}
\newcommand{\IM}{\textrm{Im}}
\newcommand{\rot}{\textrm{rot}}
\newcommand{\dv}{\boldsymbol{d}}
\newcommand{\grad}{\textrm{grad}}
\newcommand{\Tr}{\textrm{Tr}}
\newcommand{\ua}{\uparrow}
\newcommand{\da}{\downarrow}
\newcommand{\ct}{\textrm{const}}
\newcommand{\xv}{\boldsymbol{x}}
\newcommand{\mv}{\boldsymbol{m}}
\newcommand{\rv}{\boldsymbol{r}}
\newcommand{\kv}{\boldsymbol{k}}
\newcommand{\VE}{\boldsymbol{V}}
\newcommand{\sv}{\boldsymbol{s}}
\newcommand{\RV}{\boldsymbol{R}}
\newcommand{\pv}{\boldsymbol{p}}
\newcommand{\PV}{\boldsymbol{P}}
\newcommand{\EV}{\boldsymbol{E}}
\newcommand{\DV}{\boldsymbol{D}}
\newcommand{\BV}{\boldsymbol{B}}
\newcommand{\HV}{\boldsymbol{H}}
\newcommand{\MV}{\boldsymbol{M}}
\newcommand{\be}{\begin{equation}}
\newcommand{\ee}{\end{equation}}
\newcommand{\ba}{\begin{eqnarray}}
\newcommand{\ea}{\end{eqnarray}}
\newcommand{\bq}{\begin{eqnarray*}}
\newcommand{\eq}{\end{eqnarray*}}
\newcommand{\pa}{\partial}
\newcommand{\f}{\frac}
\newcommand{\FV}{\boldsymbol{F}}
\newcommand{\ve}{\boldsymbol{v}}
\newcommand{\AV}{\boldsymbol{A}}
\newcommand{\jv}{\boldsymbol{j}}
\newcommand{\LV}{\boldsymbol{L}}
\newcommand{\SV}{\boldsymbol{S}}
\newcommand{\av}{\boldsymbol{a}}
\newcommand{\qv}{\boldsymbol{q}}
\newcommand{\QV}{\boldsymbol{Q}}
\newcommand{\ev}{\boldsymbol{e}}
\newcommand{\uv}{\boldsymbol{u}}
\newcommand{\KV}{\boldsymbol{K}}
\newcommand{\ro}{\boldsymbol{\rho}}
\newcommand{\si}{\boldsymbol{\sigma}}
\newcommand{\thv}{\boldsymbol{\theta}}
\newcommand{\bv}{\boldsymbol{b}}
\newcommand{\JV}{\boldsymbol{J}}
\newcommand{\nv}{\boldsymbol{n}}
\newcommand{\lv}{\boldsymbol{l}}
\newcommand{\om}{\boldsymbol{\omega}}
\newcommand{\Om}{\boldsymbol{\Omega}}
\newcommand{\Piv}{\boldsymbol{\Pi}}
\newcommand{\UV}{\boldsymbol{U}}
\newcommand{\iv}{\boldsymbol{i}}
\newcommand{\nuv}{\boldsymbol{\nu}}
\newcommand{\muv}{\boldsymbol{\mu}}
\newcommand{\lm}{\boldsymbol{\lambda}}
\newcommand{\Lm}{\boldsymbol{\Lambda}}
\newcommand{\opsi}{\overline{\psi}}
\renewcommand{\tan}{\textrm{tg}}
\renewcommand{\cot}{\textrm{ctg}}
\renewcommand{\sinh}{\textrm{sh}}
\renewcommand{\cosh}{\textrm{ch}}
\renewcommand{\tanh}{\textrm{th}}
\renewcommand{\coth}{\textrm{cth}}

\begin{abstract}
A modified version of the Reissner-Nordstrom metric is proposed on the grounds of the nonlinear electrodynamics model. The source of curvature is an anisotropic fluid with $p_{r} = -\rho$ which resembles the Maxwell stress tensor at $r >> q^{2}/2m$, where $q$ and $m$ are the mass and charge of the particle, respectively. We found the black hole horizon entropy obeys the relation $S = |W|/2T = A_{H}/4$, with $W$ the Komar energy and $A_{H}$ the horizon area. The electric field around the source depends not only on its charge but also on its mass. The corresponding electrostatic potential $\Phi(r)$ is finite everywhere, vanishes at the origin and at $r = q^{2}/6m$ and is nonzero asymptotically, with $\Phi_{\infty} = 3m/2q$.
 \end{abstract}
 
 \section{Introduction}
 To avoid the black hole (BH) singularity problem, a lot of regular models have been proposed \cite{JB, BS, KB, AB, BF, MPS, BG1, BR, NSS, MKP, SAH}. Bardeen \cite{JB} was the first author presenting a regular BH model. However, the physical source associated to his solution was clarified much later, when Ayon-Beato and Garcia \cite{BG2} interpreted it as the gravitational field of a nonlinear magnetic monopole of a self-gravitating magnetic field.
 
 In the framework of general relativity (GR) one can find singularity-free solutions of the Einstein field equations coupled to nonlinear electrodynamics which in the weak-field approximation becomes the linear Maxwell field. The electric field is bound everywhere and asymptotically behaves as a Coulomb field. As far as the geometry is concerned, its asymptotic behavior is that of Reissner-Nordstrom (RN) type. Bronnikov and Shikin \cite{BS} and Bronnikov \cite{KB} showed that, in the Maxwell weak-field limit (small electromagnetic field scalar $F = F^{ab}F_{ab}$), static spherically symmetric general relativistic configurations with a nonzero electric charge cannot have a regular center. Nevertheless, they did not specified what does ``small $F$'' exactly mean. Therefore we consider their conclusion questionable. A counterexample will be given by our metric (2.5). 
 
  The authors of \cite{BG2} proposed a new form for nonlinear electrodynamics which coupled to gravity gives rise to a nonsingular BH solution obeying the weak energy condition. The Einstein - nonlinear electrodynamics field equations are derived from an action within the Einstein - dual nonlinear electrodynamics \cite{BG3}. Recently Balart and Vagenas \cite{BV1} built a static, charged, regular BH in the framework of Einstein - nonlinear electrodynamics theory, satisfying the weak energy condition. Their action of GR coupled to nonlinear electrodynamics can be written as \cite{BV2}
  \begin{equation}
  S = \int{\left(\frac{R}{16\pi} - \frac{1}{4\pi} L(F)\right)\sqrt{-g}d^{4}x}   
 \label{1.1}
 \end{equation}
where the Lagrangean $L(F)$ is a nonlinear function of the electromagnetic scalar $F = (1/4)F^{ab}F_{ab}$, which describes Maxwell theory for weak fields. They further studied regular charged BH metrics by means of mass distribution function and constructed the corresponding electric field for each BH solution in terms of continuous probability distributions.

We shall employ in this paper a modified version of Schwarzschild (KS) metric \cite{HC} to obtain a regular charged BH. We already noticed in \cite{HC} that the anisotropic stress tensor is similar to the Maxwell stress tensor far from a point charge located at the origin. Taking a different value for the constant $k$ and giving it another meaning, we extend the validity of the anisotropic energy-momentum tensor to all distances, even close to the electric source. We also found that the electrostatic potential derived from the corresponding electric field has a nonzero value at infinity, $\Phi_{\infty} = 3mc^{2}/2q$ where $m$ and $q$ are the source mass and charge, respectively, and $c$ is the velocity of light.

Henceforth we are employing the geometric units $G = c = 1$.

\section{Regular Reissner-Nordstrom metric}
To begin with, we consider firstly the Xiang \textit{et al.} \cite{XLS} modified form of the KS standard metric
   \begin{equation}
  ds^{2} = -(1 + 2\Psi) dt^{2} + (1 + 2\Psi)^{-1} dr^{2} + r^{2} d \Omega^{2}, 
 \label{2.1}
 \end{equation}
 where $\Psi(r) = -(m/r) exp(-\epsilon(r))$ (with $\epsilon(r) > 0)$ and $d \Omega^{2}$ stands for the metric on the unit 2-sphere. The unknown function $\epsilon(r)$ is a damping factor needed to remove the singularity when $r \rightarrow 0$ \cite{XLS}. 
  Xiang \textit{et al.} used the approximation $e^{\epsilon} > 1 + \epsilon$ to obtain $m = (\xi/2)e^{\epsilon(\xi)} > \xi \sqrt{\epsilon}$, which is valid for any $\xi$ (the horizon radius) and, therefore, they concluded that $\epsilon \propto 1/r^{2}$. Let us go to the next term in the power series development of $e^{\epsilon}$ and write
   \begin{equation}
e^{\epsilon} > 1 + \epsilon + \frac{\epsilon^{2}}{2},
 \label{2.2}
 \end{equation}
whence, with the help of the average formula
   \begin{equation}
 m = \frac{\xi}{2} e^{\epsilon(\xi)} > \frac{1}{3}\frac{3\xi}{2} (1 + \epsilon + \frac{\epsilon^{2}}{2}) > \sqrt[3]{\frac{3\xi}{2} \frac{3\xi \epsilon}{2} \frac{3\xi \epsilon^{2}}{4}} =  \frac{3}{2\sqrt[3]{2}}\xi \epsilon. 
  \label{2.3}
 \end{equation}
According to the Xiang \textit{et al.} prescription, the last term of (2.3) should be a constant, \textit{i.e.} $\epsilon \propto 1/r$.

 Our main assumption, therefore,  is to consider $\Psi(r)$ to be of the form \footnote{Let us mention that recently Ghosh \cite{SG} used the exponential cutoff $exp(-k/r)$ from (2.4) to regulate the rotating black hole solution. Nevertheless, his statement at p.11 that the exponential regulator factor from (2.4) was suggested by Brown \cite{MB} is not valid, in our view. Brown proposed the exponential convergence factor $exp(-l_{P}^{2}/(x - y)^{2})~ ((x - y)^{2}$ is the distance squared between any two points in the background spacetime) and not our exponential factor. Therefore, Brown's factor is closer to the Xiang et al. one, i.e. $e^{-\epsilon(r)}$, with $\epsilon(r) \propto 1/r^{2}$. Hence, to our knowledge, the regulator factor $e^{-\frac{k}{r}}$ has been firstly introduced in \cite{HC}}.
    \begin{equation}
    \Psi(r) = - \frac{m}{r} e^{-\frac{k}{r}}
 \label{2.4}
 \end{equation}
  where $k$ is a positive constant. With this expression for $\Psi(r)$, our proposed modified metric can be written as 
     \begin{equation}
   ds^{2} = -\left(1 - \frac{2m}{r} e^{-\frac{k}{r}}\right) dt^{2} + \frac{1}{1 - \frac{2m}{r} e^{-\frac{k}{r}}} dr^{2} + r^{2} d \Omega^{2},         
 \label{2.5}
 \end{equation}
 Being interested to relate $k$ with the charge $q$ associated to the mass $m$, we choose it to be 
     \begin{equation}
     k = \frac{q^{2}}{2m}
 \label{2.6}
 \end{equation}
 and the metric function $f(r) \equiv 1 + 2\Psi(r)$ becomes
 \begin{equation}
  f(r) =  1 - \frac{2m}{r} e^{-\frac{q^{2}}{2mr}}.   
 \label{2.7}
 \end{equation}
It should be noted that, asymptotically, the metric function tends to its RN value, i.e. $f(r) =  1 - 2m/r + q^{2}/r^{2}$ if one preserves only the first term in the power series development of the exponential function. The first derivative of $f(r)$ can be written as
 \begin{equation}
  f'(r) = \frac{2m}{r^{2}}\left(1 - \frac{q^{2}}{2mr}\right)e^{-\frac{q^{2}}{2mr}}, 
 \label{2.8}
 \end{equation}
 which has a root at $r = q^{2}/2m$. The metric function acquires its minimal value at $r = \frac{q^{2}}{2m}$ and we have
  \begin{equation}
  f_{min} = f(k) = f(\frac{q^{2}}{2m}) = 1 - \left(\frac{2m}{q\sqrt{e}}\right)^{2}
 \label{2.9}
 \end{equation}
  We distinguish three situations of interest here (Fig.1):\\
 (1) $|q| < 2m/\sqrt{e}$, when $f(k) < 0$. Equation $f(r) = 0$ has two roots, $r_{-} < q^{2}/2m$ (the Cauchy inner horizon) and $r_{+} > q^{2}/2m$ (the event horizon). However, their position cannot be determined exactly because $f(r) = 0$ is a transcendent equation. \\
 (2) $|q| = 2m/\sqrt{e}$, which leads to $f(k) = 0$. We have now a double root at $r_{H} = q^{2}/2m = 2m/e$ which represents the event horizon. Therefore, $f(r_{H}) = 0$ and $f'(r_{H}) = 0$ are simultaneously satisfied. This looks like an extremal BH, with a degenerate horizon \cite{BR, HC}\\
 (3) $|q| > 2m/\sqrt{e}$, \textit{i.e.} $f(k) > 0$. The equation $f(r) = 0$ has no roots in this case. The metric function is positive for any $r$. For the case (1) the authors of \cite{BV2} give the expressions of $r_{+}$ and $r_{-}$ in terms of the Lambert W-function. The case (2) has been analyzed in detail in Ref. \cite{HC}. We consider from now on that there are two horizons (case (1)) and that $q > 0$. 
 
 Let us take now a static observer with the velocity field
  \begin{equation}
   u^{b} = \left(\frac{1} {\sqrt{1 - \frac{2m}{r} e^{-\frac{q^{2}}{2mr}}}}, 0, 0, 0\right)    
 \label{2.10}
 \end{equation} 
  where $b$ labels $(t,r,\theta,\phi)$. The acceleration 4 - vector  $a^{b} = u^{a}\nabla_{a}u^{b}$  is given by
  \begin{equation}
  a^{b} = \left(0, \frac{m(1 - \frac{q^{2}}{2mr})} {r^{2}} e^{-\frac{q^{2}}{2mr}}, 0, 0, \right)    
 \label{2.11}
 \end{equation} 
 and we note that $a^{r}$ is vanishing when $r \rightarrow 0$ and at $r = q^{2}/2m $, where the metric function $f(r)$ reaches its lowest value. In addition, the gravitational field becomes repulsive for $r < q^{2}/2m$, when $a^{r} < 0$. 
 
 Once we have localized the event horizon $H$ at $r_{+}$, where $f(r_{+}) = 0$, the surface gravity may be written as
   \begin{equation}
 \kappa =  \sqrt{ a^{b}a_{b}} \sqrt{-g_{tt}}|_{H} = \frac{1}{2r_{+}}\left(1 - \frac{q^{2}}{2mr_{+}}\right),
 \label{2.12}
 \end{equation} 
 which vanishes when $r_{+} = q^{2}/2m $, i.e. for the extremal BH \cite{HC}.
 
 \section{Anisotropic energy-momentum tensor}
 We look now for the sources of the spacetime (2.5), namely the stress tensor to lie on the r.h.s. of Einsteins' equations $G_{ab} = 8\pi T_{ab}$ in order that (2.5) to be an exact solution. 
 By means of the software package Maple - GRTensorII, one finds that
   \begin{equation}
   \begin{split}
  T^{t}_{~t} = -\rho = -\frac{mk}{4\pi r^{4}} e^{-\frac{k}{r}},~~~ T^{r}_{~r} = p_{r} = - \rho,\\ T^{\theta}_{~\theta} = T^{\phi}_{~\phi} = p_{\theta} = p_{\phi} =  \frac{ mk}{8 \pi r^{5}} (2r - k) e^{-\frac{k}{r}}.
  \end{split}
\label{3.1}
\end{equation}
 When $k = 2m/e$ (where ``$e$'' is the Euler's number) we retrieve the components of the stress tensor from \cite{HC}, i.e. the extremal BH. The case $k =q^{2}/2m $ leads to the Balart-Vagenas stress tensor \cite{BV2}. Their mass $M$ corresponds to our $m$ and $\sigma(r)$, related to the mass function, is proportional to the Misner-Sharp mass $m(r)$ obtained from
  \begin{equation}
  1 - \frac{2m(r)}{r} = g^{ab}\nabla_{a}r \nabla_{b}r,  
 \label{3.2}
 \end{equation} 
which in our situation becomes
  \begin{equation}
  m(r) = m e^{-\frac{q^{2}}{2mr}}.
 \label{3.3}
 \end{equation} 
 From (3.1) we notice that $\rho > p_{\theta}$ always and $p_{r} = - \rho$, as for dark energy.  Nevertheless, the fluid is anisotropic since $p_{r} \neq p_{\theta} = p_{\phi}$. The energy density and all pressures are non-singular at $r = 0$ and when $r \rightarrow \infty$ (where, actually, they vanish). Moreover, $\rho$ is positive for any $r$ and the weak energy condition is fulfilled. However, the strong energy condition is not satisfied for $r < k/2$, where $\rho + \Sigma p_{i} = 2p_{\theta} < 0 ~(i = 1,2,3)$. For $r >> k$ and with $k = q^{2}/2m$, $\rho(r)$ no longer depends on $m$ but only on $q$. In addition, we have $\rho \propto (q/r^{2})^{2}$, i.e. it is proportional to the Coulomb field squared. Moreover, again for $r >> q^{2}/2m$ all the components of (3.1) acquires exactly the form of the Maxwell stress tensor
  \begin{equation}
 T^{a}_{(e)b} = \frac{q^{2}}{8\pi r^{4}} (-1, -1, 1, 1),
 \label{3.4}
 \end{equation} 
representing the static electric field of a point charge of strength $q$ and mass $m$, located at $r = 0$.

We conjecture therefore that the stress tensor (3.1) is valid at any $r > 0$, both in gravity and electrostatics, with $k$ chosen accordingly. When $k = 2m/e$ we get the extremal BH form \cite{HC} and KS spacetime is obtained asymptotically. When $k = q^{2}/2m$, $T^{a}_{~b}$ depends both on $q$ and $m$ and gravity is mixed with nonlinear electrodynamics. The classical electrostatics is recovered at $r >> q^{2}/2m$, when (3.1) becomes exactly the Maxwell stress tensor (3.4) and the RN metric is obtained. One should also note that all the physical parameters from (3.1) vanish both at $r = 0$ and at infinity. It is worth noting that if $k \propto q$, the cutoff factor would be $exp(-q\sqrt{G}/4\pi \epsilon_{0}c^{2}r) \approx exp(-q/10^{12}r)$, which shows that we need a huge $q/r$ for the exponential factor to play some role.

\section{Energetic considerations}
Our first task in this chapter is to evaluate the gravitational Komar energy of the spacetime (2.5), with $k = q^{2}/2m$
   \begin{equation}
W = 2 \int(T_{ab} - \frac{1}{2} g_{ab}T^{c}_{~c})u^{a} u^{b} N\sqrt{\gamma} d^{3}x ,
\label{4.1}
\end{equation}
which is measured by a static observer with the velocity field $u^{a}$. $N$ in (4.1) is the lapse function and $\gamma$ is the determinant of the spatial 3-metric. The above equation yields
 \begin{equation}
 W(r) = \int^{r}_{0}\frac{q^{2}}{r'^{4}}\left(1 - \frac{q^{2}}{4mr'}\right)e^{-\frac{q^{2}}{2mr'}}r'^{2}dr'.  
\label{4.2}
\end{equation}
The substitution $x = 1/r'$ leads finally to
 \begin{equation}
 W(r) = m\left(1 - \frac{q^{2}}{2mr}\right) e^{-\frac{q^{2}}{2mr}}
\label{4.3}
\end{equation}
($W$ is not, of course, defined at $r' = 0$ and therefore we evaluated the limit $r' \rightarrow 0$, which is zero). 

 It is clear from (4.3) that $W = m$ at infinity, as expected. In other words, the total Komar energy of the anisotropic fluid equals the rest mass of the central source at infinity. Nevertheless, $W$ becomes negative for $r < q^{2}/2m$ and presents a minimum $W_{min} = -m/e^{2}$ (which does not depend on $q$, as in \cite{HC}) at $r = q^{2}/4m$ (Fig.2). This seems to be a steady state,  rooted from the negative pressures contribution. 
 
  Let us compute now the horizon entropy $S = |W|/2T$ \cite{TP}, where $T = \kappa/2\pi$ is obtained from the surface gravity (2.12). With $r_{+}$ from $f(r_{+}) = 0$, one obtains
  \begin{equation}
 S_{H} = \frac{m\left(1 - \frac{q^{2}}{2mr_{+}}\right) e^{-\frac{q^{2}}{2mr_{+}}}}{\frac{1}{2\pi r_{+}}\left(1 - \frac{q^{2}}{2mr_{+}}\right)} = \pi r_{+}^{2},
\label{4.4}
\end{equation}
 i.e. the relation $S_{H} = A_{H}/4$ is obtained, as it should be for a BH.
 
 We wish now to find the expression of the electrostatic potential energy in the framework of the nonlinear electrodynamics. To reach that goal we will make use of the expression of the electric field $E = F_{tr}$ obtained by Balart and Vagenas in \cite{BV2} 
  \begin{equation}
 E(r) = \frac{q}{r^{2}} \left(1 - \frac{q^{2}}{8mr}\right) e^{-\frac{q^{2}}{2mr}}
\label{4.5}
\end{equation}
where $F_{tr} = - F_{rt}$ stands for the only nonzero component of the electromagnetic field tensor. On the grounds of their nonlinear electrodynamic model, Eq. (4.5) tells us that $E$ depends not only on $q$ but also on $m$. That means $E$ changes when $q$ is fixed but $m$ is varied. However, $E$ returns to its Coulombian form when $r >> q^{2}/2m$. If we evaluate $q^{2}/2m$ for an electron, we get $k_{e} \equiv q_{e}^{2}/2m_{e}c^{2}$, which is half the electron radius $r_{e}$. As $k_{e} > 2m_{e}/e$, there are no horizons (we have $m_{e} < q_{e}\sqrt{e}/2$; this is a little bit different compared to the RN case, where no horizon means $m_{e} < q_{e}$). We would like to evaluate the electron electric field at, say, $r = r_{e} \approx 10^{-15} m$. One obtains
  \begin{equation}
 E_{e} = \frac{q_{e}}{r_{e}^{2}} \left(1 - \frac{q_{e}^{2}}{8m_{e}r_{e}}\right) e^{-\frac{q_{e}^{2}}{2m_{e}r_{e}}} \approx \frac{q_{e}}{r_{e}^{2}}(1 - \frac{1}{8})\frac{1}{\sqrt{e}} = \frac{7}{8\sqrt{e}} \frac{q_{e}}{r_{e}^{2}},
\label{4.6}
\end{equation}
 which is less than the Coulombian value $q_{e}/r_{e}^{2}$ ($m_{e}$ and $q_{e}$ are, respectively, the electron mass and charge). Let us take now a sphere of mass $m = 10^{12} Kg$, radius $R = 1 Km$ and charged with $q = 1C$. We have $q^{2}/2m \approx 5.10^{-18} cm$ and $2m/e \approx 5.10^{-14} cm$, so that $q^{2}/8\pi \epsilon_{0}mc^{2} << 2Gm/ec^{2}$, if we introduce all fundamental constants. Therefore, there are horizons and 
 \begin{equation}
1 - \frac{2m}{r_{+}} e^{-\frac{q^{2}}{2mr_{+}}} \approx 1 - \frac{2m}{r_{+}} = 0    
\label{4.7}
\end{equation}
 (note that $r_{+} > 2m/e$). Hence $r_{+} \approx 2m = 10^{-13} cm$, far inside the sphere. In other words, for a macroscopic mass $m$ the nonlinear electrodynamics has a linear behavior.
 
 The electric field $E(r)$ given by (4.5) vanishes at $r = q^{2}/8m$, when $r \rightarrow 0$ or $r \rightarrow \infty$. To find how $E(r)$ varies with $r$ we need to solve the equation 
   \begin{equation}
 E'(r) = -\frac{2q}{r^{3}} \left(1 - \frac{7}{8} \frac{q^{2}}{2mr} +\frac{1}{8}\left(\frac{q^{2}}{2mr} \right)^{2}\right) e^{-\frac{q^{2}}{2mr}} = 0
\label{4.8}
\end{equation}
 which yields
 \begin{equation}
 r_{1} = \frac{q^{2}}{32m}(7 - \sqrt{17}),~~~r_{2} =  \frac{q^{2}}{32m}(7 + \sqrt{17})  
\label{4.9}
\end{equation}
The plot of $E(r)$ versus $r$ is represented in Fig.3 ($r_{1}$ and $r_{2}$ are located at the same distance w.r.t. $q^{2}/8m$).

It is worth to find the electrostatic potential $\Phi(r)$ where the electric field (4.5) is rooted from
 \begin{equation}
 \Phi(r) = -\int{E(r)dr}
\label{4.10}
\end{equation} 
 Eq. (4.10) gives us 
  \begin{equation}
 \Phi(r) = - \int^{r}{\frac{q}{r'^{2}} \left(1 - \frac{q^{2}}{8mr'}\right) e^{-\frac{q^{2}}{2mr'}}dr'}
\label{4.11}
\end{equation} 
After a separation of (4.11) in two integrals and the substitution $x = q^{2}/2mr'$ we arrive at
  \begin{equation}
 \Phi(r) = \frac{3m}{2q} \left(1 - \frac{q^{2}}{6mr}\right) e^{-\frac{q^{2}}{2mr}} + C,
\label{4.12}
\end{equation}
where $C$ is a constant of integration. In order for $\Phi(r)$ to vanish at $r = 0$, we choose $C = 0$. However, $\Phi(r)$ tends to $\Phi_{\infty} = 3m/2q$ when $r \rightarrow \infty$ (Fig.4). That means the electric potential energy is $q \Phi_{\infty} = 3m/2$. This resembles the gravitational potential $\Phi_{g} = -g_{tt} = 1 - 2m/r$ of the KS spacetime which may be written as $\Phi_{g} = c^{2} - 2Gm/r$ if we introduce the fundamental constants $G$ and $c$. One obtains $m\Phi_{g,\infty} = mc^{2}$ and the rest energy appears here as the gravitational potential energy \cite{HC1}. It is interesting to observe that $\Phi(r)$ from (4.12) depends not only on $q$ and $m$ but also on the speed of light. 

 In the limiting case $r >> q^{2}/2m$, a power series development yields
   \begin{equation}
 \Phi(r) \approx \frac{3m}{2q} -\frac{q}{r} \left(1 - \frac{5}{8} \frac{q^{2}}{2mr} +\frac{1}{4}\left(\frac{q^{2}}{2mr} \right)^{2} - \frac{1}{16} \left(\frac{q^{2}}{2mr} \right)^{3} + ...\right)
\label{4.13}
\end{equation}
Apart from the first constant term on the r.h.s. of (4.13), we recover the Coulombian term $-q/r$ which is the only term that does not depend on the mass of the spherical source and the speed  of light. Moreover, if $q$ was negative, $\Phi(r)$ should change sign, just like $E(r)$. On the contrary, $W(r)$ is insensitive at that change.

\section{Conclusions}
Our purpose in this paper was to find singularity-free solutions of Einstein's equations coupled to nonlinear electrodynamics. The asymptotic behavior of our solution is that of the RN type and the Maxwell stress tensor for a static charge is recovered as the source of the gravitational field equations. The regular charged BH is obtained from a modified version of the KS metric, giving a different meaning for the constant $k$ in terms of the charge of the source. 

Our stress tensor (3.1) was retrieved from \cite{HC} and has been also recently computed by Balart and Vagenas using probability distribution function. By means of their expression for the electric field $E(r)$, we deduced the corresponding electrostatic potential $\Phi(r)$ which vanishes at $r = 0$ but tends to the value $3m/2q$ asymptotically. A power series development of $\Phi(r)$ for $r>>q^{2}/2m$ contains the Coulombian potential $-q/r$ as the only term that does not depend on the source mass and the velocity of light.\\
\textit{Note added in proof}. After this paper has been completed, we were aware of the preprint \cite{PG}, where the author postulated a universe of uniform charge density at a large scale which induces an electrostatic potential similar with our $\Phi_{\infty} = 3m/2q$.

\begin{figure}
\includegraphics[width=12.0cm]{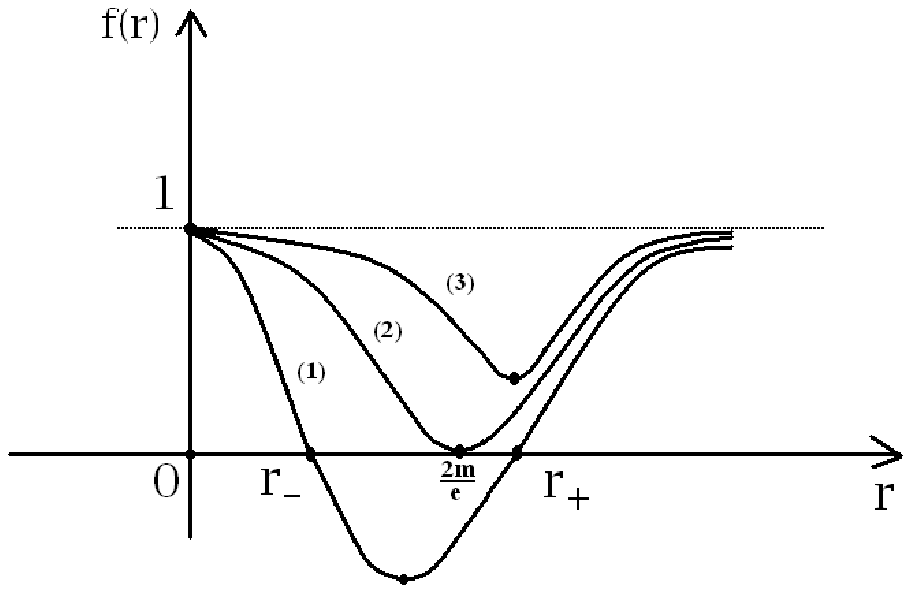}
\caption{ The metric function $f(r)$ against $r$: (1)~ $|q|<2m/\sqrt{e} ~(when ~f(k)<0); (2)~ |q|=2m/\sqrt{e}  ~(when ~f(k) = 0); (3)~|q|>2m/\sqrt{e} ~ (when ~f(k)>0)$, where $k$ is a positive constant. At $r_{-}$ and $r_{+}$ are located the inner Cauchy horizon and the event horizon, respectively.}  
\label{fig1}
 \includegraphics[width=12.0cm]{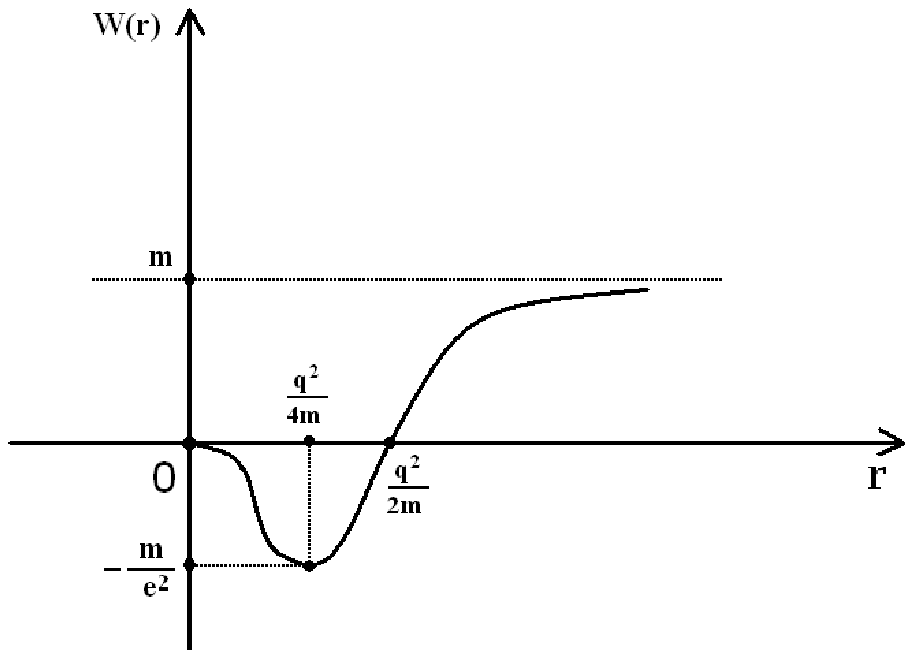}
\caption{The Komar energy versus $r$. $W(r)$ reaches its minimal value $-m/e^{2}$ at $r = q^{2}/4m$ and vanishes at $r = q^{2}/2m$.}
\label{fig2}
\end{figure}
\begin{figure}
\includegraphics[width=12.0cm]{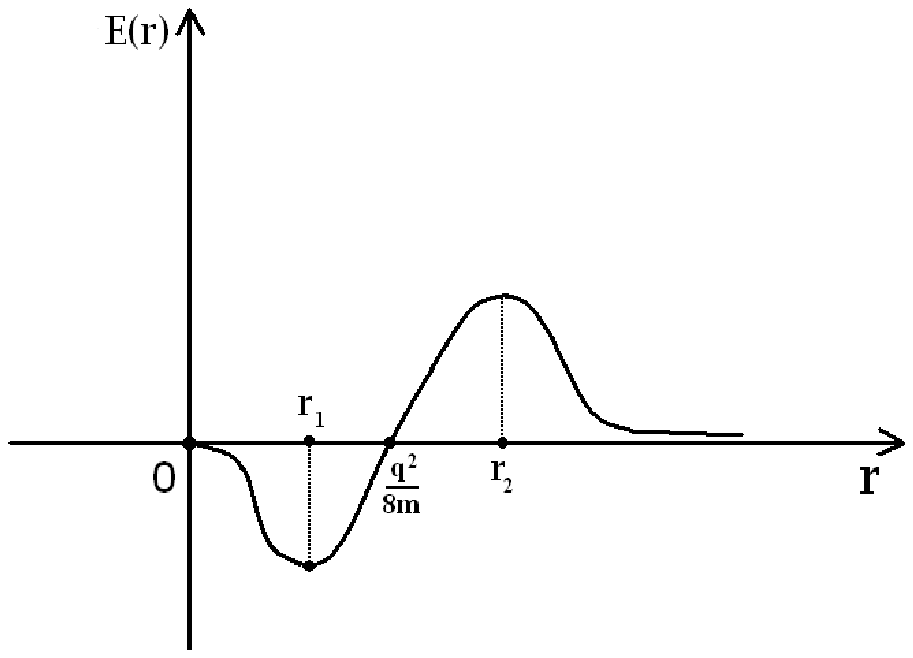}
\caption{The electric field in nonlinear electrodynamics. $E(r)$ vanishes at $r = q^{2}/8m$ and reaches its extremal values at $r_{2,1} = q^{2}(7 \pm \sqrt{17})/32m$.}
\label{fig3}
\includegraphics[width=12.0cm]{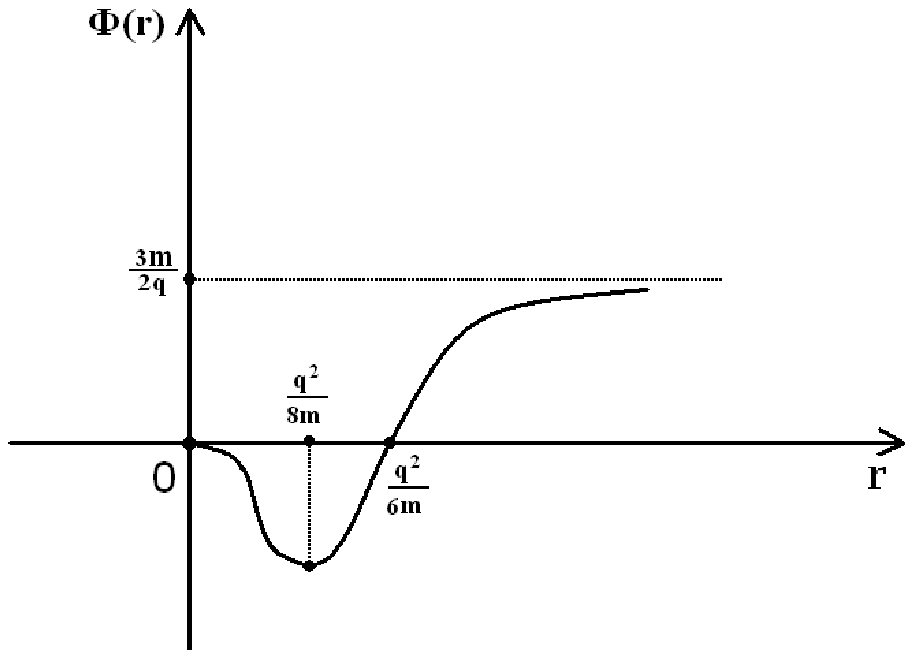}
\caption{The plot of the electric potential as a function of $r$. $\Phi(r)$ has a minimum at $r = q^{2}/8m$, vanishes at $r = q^{2}/6m$ and tends to $3m/2q$ asymptotically.}
\label{fig4}
\end{figure}

\end{document}